\magnification=1200
\headline={\ifnum\pageno=1 \hfil \else \hfil \tenrm \folio \hfil \fi}
\footline={\hfil}
\null \vskip10mm 
\centerline{\bf SCALE INVARIANT COSMOLOGY}
\quad \par
\quad \par
\quad \par
\centerline{{\bf Pierre MIDY} \footnote{${\rm \sp{*}}$}{e-mail: Pierre.Midy@cri.u-psud.fr}} 
\quad \par
\centerline{\bf C.R.I., Universit\'e Paris XI}
\centerline{\bf Centre d'Orsay, F 91405 Orsay}
\quad \par
\centerline{\bf and}
\quad \par
\centerline{{\bf Jean-Pierre PETIT} \footnote{${\rm \sp{**}}$}{For scientific correspondence: Villa Jean-Christophe, Chemin de la Montagn\`ere, \break F 84120 Pertuis
\ -- \ e-mail: perso@jp-petit.com}}
\quad \par
\centerline{\bf Observatoire de Marseille}
\centerline{\bf 2, Place Le Verrier, F 13248 Marseille Cedex 4}
\quad \par
\quad \par
\quad \par
\quad \par
\centerline{\bf Abstract}
\quad \par
\quad \par
An attempt is made here to extend to the microscopic domain the scale 
invariant character of gravitation -- which amounts to
consider expansion as applying to any physical scale. Surprisingly, 
this hypothesis does not prevent the redshift from being obtained. It leads to
strong restrictions concerning the choice between the presently available 
cosmological models and to new considerations about the notion of time. 
Moreover, there is no horizon problem and resorting to inflation is not
necessary. \par
\vfill \eject
\indent {\bf 1.\ \ Introduction} \par
\vskip5mm 
Since Hubble's discovery of the recession of galaxies, obtaining the variation
with cosmic time of the scale parameter (or "radius") $R(t)$ which describes the
expansion of the universe is the basic problem a cosmological theory has to 
deal with. The closely related question of deciding relative to what the
expansion takes place, or where it stops, is much more seldom explicitly 
stated [1]. It is commonly admitted there is no expansion at distances smaller
than the size of clusters of galaxies, so that this size, or the size of
our galaxy, or of the solar system, \dots\  can be used as a reference scale.
This may seem to be sensible at present, but what about the past, when $R(t)$ was smaller than these objects? The existence and observability of permanent 
reference scales for time and length intervals thus appear to be assumed. 
It is also commonly admitted -- but this argument has been questioned [2, 3] -- that 
the scale parameter cannot be applied to all length scales in the universe, on 
the grounds that this would amount to getting no expansion at all, so that the universe would be scale invariant and there would be no redshift. On the other 
hand, if the inflation theory eliminates the causality problem for the 
presently observed regions of the universe, it does not in the future \dots\  
In order to solve these problems, L. Nottale [1] introduces a transition from 
scale independence to scale dependence in the general frame of a theory of
scale relativity. \par 
Already Laplace [1, 4] had pointed out the scale invariance of Newton's theory of
gravitation, concluding that "the universe reduced to the smallest imaginable 
space would always present the same appearance to observers" and that "the laws
of nature only permit us to observe relative dimensions".
Poincar\'e made similar observations as concerns comparing time
intervals as well as lengths at two different instants (and this is, of course,
the only way to proceed for time intervals): he notes "we have no direct 
intuition of the equality of two lapses of time" -- for instance those "between
noon and one o'clock and between two and three o'clock" [5]. So it should be
possible to allow a general variation of all length [2] and time intervals
as long as one keeps in mind that this kind of variation cannot be physically 
observed by comparisons necessarily performed at the same instant: there would
be no permanent observable reference scale; since this
looks as an unnecessary complication, the possibility for expansion to act at
any scale has seldom been considered seriously. However, Hoyle and Narlikar 
[6], as an alternative to expansion, 
already used a somewhat similar view in their variable mass theory, where all 
masses are supposed to increase with cosmic time, which implies decreasing 
wavelengths for atomic radiations and decreasing sizes for atoms. 
Indeed, it may be observed that, as opposed to the case
of pure gravitation, where the size and the period of a Keplerian orbit can be
changed arbitrarily as long as Kepler's laws are respected, without any
modification of Newton's constant nor of the masses of the bodies, physical
constants must be allowed to change if the expansion is considered to apply,
for instance, to structures the size of which is directly linked to atomic 
sizes. This is clear from the expression of the Bohr 
radius, 
 $$ a \sb{0}\ =\ {h \sp{2} \epsilon \sb{0} \over \pi m \sb{e} e \sp{2}} \ =\ {\hbar \over m_e \alpha c} \quad , \eqno (1)$$
which involves physical constants only: supposing the absence of 
permanent observable reference scales for length and time intervals implies 
some physical constants, among which $c$, can vary with cosmic time. \par
So, since $c$ also appears in the expression of the metric tensor, we shall
suppose it to be variable and we shall look for a function $R(t)$ which already 
resulted from one of the 
classical cosmological theories: such a function should not depend on
hypotheses which can be made about the reference scale relative to which $R(t)$ 
is defined. This leads to a highly constrained problem which cannot be solved
realistically (i.e. with non-zero pressure and density) unless more than four
dimensions are introduced in the description of space-time.
 This idea, initially due to Einstein and recently reintroduced by P. Wesson
and his co-workers [7, 8], aims at obtaining the properties of matter in our 
four dimensional 
space-time in a purely geometrical way, starting from Einstein's equations
 $$ R \sb{\mu \nu}\ =\ 0 \eqno (2) $$
written for five dimensions or more with no matter-energy terms on the  
right-hand side to find the usual 4-dimensional equations 
 $$ R \sb{\mu \nu} \sp{(1+3)}\ =\ - {8 \pi G \over c^2} S \sb{\mu \nu}\ =\ 
- {8 \pi G \over c^2}\, \bigl(T_{\mu \nu} - {1 \over 2} g_{\mu \nu} 
T^{\lambda}_{\ \lambda}\bigr) \eqno (2') $$
where $T_{\mu \nu}$ is the energy-momentum tensor of a perfect fluid, and we 
shall try it together with the hypothesis of a variable $c$.\par
\vskip10mm
\indent {\bf 2.\ \ The Einstein equations}\par
\vskip5mm
 We write Einstein's equations for an arbitrary number $n$ of space dimensions  and with a variable $c$ in a maximally symmetric space, hence with a metric 
which, as concerns space, generalizes to $n$ dimensions the metric of 
Robertson-Walker. Choosing 
such a metric with $n$ greater than 3 is obviously surprising -- usually a 
different
scale factor is associated with space dimensions beyond the third one -- and 
this will be discussed later (\S\kern.15em \S\kern.15em 5 and 6). \par
 It is to be noted that, as a consequence of our hypotheses, the geometrical space-time coordinates -- defined with respect to
permanent reference scales -- will appear to be distinct from the physically observable ones. This is obvious for lengths:
expansion applying to spatial domains of arbitrary size, physical distances here are the comobile ones; the case of time intervals
will be dealt with in \S\kern.15em \S\kern.15em 3 and 4. So as to make this distinction easier, geometrical coordinates or intervals will generally
be written using Latin characters, and physical coordinates or intervals using Greek ones. \par 

\let\\=\partial
 Designing by $0$ the index corresponding to the chronological coordinate $x^0$, with
$dx^0 = c(t) dt$, and by
$i, j = 1, 2,\dots , n$ the indices associated with the spatial coordinates, the
components of the metric tensor are: 
 $$g_{0 0}\ =\   -1 \eqno (3) $$
 $$g_{i 0}\ =\   0 \eqno (4) $$
 $$g_{i j}\ =\   R^2(t) \tilde g_{i j}, \quad \tilde g_{i j}\ =\   0 \quad (i \not = j)\eqno 
(5) $$
where the $\tilde g_{i j}$ depend on the space coordinates only. Choosing $x^0$ rather
than $t$ and taking into account the time dependence of $c$ at the end of the calculations allow one to get a constant $g_{0 0}$ and thus a smaller number of non-zero components for the affine connection. Starting from
$$\Gamma^{\beta}_{\gamma \alpha}\ =\   {1\over2}g^{\beta \eta}\ [\\_{\gamma} g_{\alpha \eta} + \\_{\alpha} g_{\eta \gamma} - \\_{\eta} g_{\gamma \alpha}]
 \eqno (6) $$
(where the four indices run from $0$ to $n$),
these components are found to be (the $'$ designing a derivation with respect to
 $x^0$): 
 $$\Gamma^0_{k i}\ =\   R R' \tilde g_{k i} \eqno (7) $$
 $$\Gamma^j_{0 i}\ =\   \Gamma^j_{i 0}\ =\   {R' \over R} \delta^j_i \eqno (8) $$
 $$\Gamma^j_{k i}\ =\   \tilde \Gamma^j_{k i} \quad , \eqno (9) $$
so the $\Gamma^j_{k i}$ are time-independent. \par
 This yields for the Ricci tensor 
 $$R_{\alpha \beta}\ =\   \\_{\beta} \Gamma^{\eta}_{\eta \alpha} - \\_{\eta} \Gamma^{\eta}_{\alpha \beta} + \Gamma^{\lambda}_{\alpha \eta} \Gamma^{\eta}_{\beta \lambda} - \Gamma^{\eta}_{\eta \lambda} \Gamma^{\lambda}_{\alpha \beta} \quad , \eqno (10) $$
the non-vanishing components
 $$R_{0 0}\ =\   n {R'' \over R} \eqno (11) $$
 $$R_{i j}\ =\   \tilde R_{i j} - (R R'' + (n - 1)R'^2)\ \tilde g_{i j} \eqno (12)$$
where [9], $k = 0, \pm 1$ being the curvature index,
 $$\tilde R_{i j}\ =\   - (n - 1)k \tilde g_{i j} \quad , \eqno (13) $$
hence
 $$R_{i j}\ =\   - [R R'' + (n - 1) R'^2 + (n - 1) k]\ \tilde g_{i j} \quad .  \eqno (14)$$
 \indent Taking into account the hypothesis of a variable $c$ and introducing the time variable $t$ gives:
 $$R'\ =\   {d R \over d x^0} \eqno (15) $$
and 
 $$R''\ =\   {d^2 R \over {d x^0}^2}\ =\   {1 \over c^2} {d^2 R \over d t^2} - {1 \over c^3} {d c \over d t} {d R \over d t} \quad, \eqno (16)$$
so that (11) and (14) become
 $$R_{0 0}\ =\   {n \over R c^2}\ (R'' - {c' \over c} R') \eqno (17)$$ 
and
 $$R_{i j}\ =\   - \Bigl[{R \over c^2} (R'' - {c' \over c} R') + (n - 1)({R'^2 \over c^2} + k)\Bigr]\ \tilde g_{i j} \quad , \eqno (18)$$
where the prime now stands for the derivation relative to $t$. \par
 Of course, the same result might have been obtained using $t$ instead of $x^0$ from the 
beginning, getting
 $$g_{t t}\ =\   - c^2(t) $$
instead of (3), a non-zero
 $$\Gamma^t_{t t}\ =\   {c' \over c} $$  
and 
 $$\Gamma^t_{k i}\ =\   {R R' \over c^2} \tilde g_{k i} $$
instead of (7). \par
 Applying (2) to (17) and (18) immediately entails
 $$k\ =\   -1  \eqno (19) $$
if the density is to be non-zero, so that the $n$-space curvature is negative, and 
 $$R'^2\ =\   c^2 \quad , \eqno (20) $$
hence
 $$R'\ =\   \epsilon c \quad (\epsilon\ =\   \pm 1), \eqno (21) $$
 $$R''\ =\   \epsilon c' \quad ,\eqno (22) $$
and 
 $${c' \over c}\ =\   {R'' \over R'} \quad . \eqno (23) $$
 Noting 
 $$R^{(F;1+3)}_{0 0}\ =\   {3 R'' \over R c^2} \eqno (24)$$
and
 $$R^{(F;1+3)}_{i j}\ =\   - {1 \over c^2} (R R'' + 2 R'^2)\ \tilde g_{i j} \quad (i, j\ =\   1, 3) \eqno (24')$$
the parts of the Ricci tensor components which correspond to the Friedmann model
with zero curvature and using equations (19) to (23), the time-time (17) and 
space-space (18) components of the Ricci tensor can be written:
 $$R_{0 0}\ =\   R^{(F;1+3)}_{0 0} - {3 \epsilon c' \over R c^2} \eqno (25) $$
 $$\quad R_{i j}\ =\   R^{(F;1+3)}_{i j} + (2 + \epsilon {R c' \over c^2})\ \tilde g_{i j} \quad (i, j\ =\   1, 3) \quad .\eqno (25') $$ 
 Identifying (2) with (2') where the source term is:
 $$S_{0 0}\ =\   {1 \over 2}(\rho + {3 p \over c^2}) \eqno (26) $$
 $$S_{i 0}\ =\   0 \eqno (27) $$
 $$S_{i j}\ =\   {1 \over 2}(\rho - {p \over c^2}) R^2 \tilde g_{i j} \eqno (28) $$
gives:
 $${4 \pi G \over c^2} (\rho + {3 p \over c^2})\ =\   - {3 \epsilon c' \over 
R c^2} \eqno (29) $$
 $${4 \pi G \over c^2} (\rho - {p \over c^2})\ =\   (2 + \epsilon {R c' \over 
c^2}) {1 \over R^2} \quad . \eqno (30) $$
 Of course, for this identification to be possible, the negative curvature  
$n$-space has to contain an Euclidean three dimensional variety: an example of 
this situation will be given later (\S\kern .15em 5). \par
 Both $R$ and $c$ depending on $t$, there must be a relation between them. 
Writing, with $K$ constant, 
 $$\phi(R, c)\ =\   K \eqno (31) $$
hence 
 $${\\ \phi \over \\ R} R' + {\\ \phi \over \\ c} c'\ =\   0 \eqno (32) $$
and, using (21),
 $${c' \over c}\ =\   - \epsilon {\\ \phi / \\ R \over \\ \phi / \\ c} \eqno (33)$$
(29) and (30) give:
 $${\rho c^2 + 3 p \over \rho c^2 - p}\ =\   {3 \delta \over 2 - \delta} ,\eqno (34) $$  
where 
 $$\delta\ =\   {R \over c} {\\ \phi / \\ R \over \\ \phi / \\ c} \eqno (35) $$
 So, for an equation of state (34) with $\delta$ a constant,
 $${d c \over d R}\ =\   - {\\ \phi / \\ R \over \\ \phi / \\ c}\ =\   - \delta {c \over R} \eqno (36) $$
and 
 $$ c R^{\delta}\ =\   K' \eqno (37) $$
or, with $\gamma\ =\   1 / \delta$,
 $$ R c^{\gamma}\ =\   K'' \eqno (38) $$
where $K'$ and $K''$ are constant. \par
 The equation of state (34) can be given the form
 $$p\ =\   ({2 \over \gamma} - 1)\,{\rho c^2 \over 3} \eqno (39) $$
and includes the two cases 
 $$p\ =\   0 \eqno (40) $$
and
 $$p\ =\   {\rho c^2 \over 3} \eqno (40') $$
of the dust universe and of the radiation era for $\gamma = 2$ and $1$ 
respectively.\par
 Whatever $\gamma$, (29) and (30) yield 
 $${8 \pi G \over 3} \rho\ =\   {R'^2 \over R^2} \quad , \eqno (41) $$
so that the density of matter $\rho$ corresponds, as results from the 
elimination procedure leading to (41), to the curvature term in the expressions 
(18) or (25') of $R_{i j}$. Introducing the Hubble constant
 $$H\ =\   {R' \over R} \quad , \eqno (42) $$
(41) writes
 $$\rho G\ =\   {3 \over 8 \pi} H^2 \eqno (43) $$
-- so the density equals the critical density. 
This gives for the density of energy $\varepsilon = \rho c^2$:
 $$\quad \varepsilon \ =\ {3 c^2 \over 8 \pi G} H^2 \quad . \eqno (43') $$
Of course, the pressure is a function of $\gamma$:
 $$p\ =\   {c^2 \over 8 \pi G}\ ({2 \over \gamma} - 1)\, H^2 \quad . \eqno (44)$$
\indent The scale parameter $R(t)$ is easily obtained deriving (38) with 
respect to time and using (23) to get
 $$R'' + {R'^2 \over \gamma R}\ =\   0 \eqno (45) $$
from which it can be seen that $\delta$ is but the deacceleration parameter $q$.
 The non-trivial solution of (45) which satisfies
 $$R(0)\ =\   0 \eqno (46) $$
and
 $$R(t_0)\ =\   R_0 \eqno (46')$$
is:
 $$\quad R(t)\ =\   R_0 \Bigl({t \over t_0} \Bigr)^{\gamma \over \gamma + 1} 
\quad , \eqno (47)$$
as in the zero curvature Friedmann model for an equation of state of
the type (39). \par
Thus for the radiation era ($\gamma = 1$),
 $$R(t)\ =\   R_0 \Bigl({t \over t_0} \Bigr)^{1/2} \eqno (48) $$
and for the dust universe ($\gamma = 2$),
 $$\quad R(t)\ =\   R_0 \Bigl({t \over t_0} \Bigr)^{2/3} \quad . \eqno (49) $$
\vskip10mm 
\indent {\bf 3.\ \ Time and the redshift}\par
\vskip5mm
 In the same way as lengths remain proportional to the cosmological scale
factor $R(t)$, intervals of cosmic time vary as $t$ itself: for instance, the
period $T = l/c$ of a radiation of wavelength $l$ varies as 
 $$T \ \approx \ {R \over c}\ \approx \ R^{\gamma + 1 \over \gamma}\ \approx \ t \eqno (50) $$
and the same kind of variation will characterize the period of any circular
motion, since its speed can be defined as a fraction of $c$, and thus varies 
as $R^{-1/\gamma}$, as also results for various examples (period of a pendulum,
of planetary motions, \dots) from the gauge relations described in 
\S\kern .15em 7: every time interval expands proportionally to cosmic time
with respect to a permanent reference scale.\par
 At a given $t$, a cosmic time interval $dt$ can but be defined as proportional
to $t$ and to the number $d\vartheta$ of periods a reference clock completes during 
this interval $dt$:
 $$dt\ =\   k\ t\ d\vartheta \eqno (51)$$
where $k$ is a constant, the dimension of which should be the inverse of a time if $\vartheta$
is to be a time also. Obviously, intervals of $\vartheta$ only (and of the associated 
proper time), not of $t$, can be
measured. \par
 Let us now state the nature of $\vartheta$ more precisely.  
 The square of the (1 + 3) space-time interval with zero space curvature writes,
following from now on the timelike convention:
 $$ds^2\ =\   c^2(t) dt^2 - dl^2 \eqno (52) $$ 
where $dl$ is the space interval, or, introducing $R(t)$, which we suppose to
have the dimension of a length:
 $$ds^2\ =\   c^2(t) dt^2 - R^2(t) d \xi ^2 \eqno (52') $$
and here $d \xi$ stands for the spatial distance in dimensionless comoving
coordinates. \par
 Computing $c(t)$ (with $\epsilon$\ =\   1) from (21) and (47):
 $$c(t)\ =\    {\gamma \over \gamma + 1}{R_0 \over t_0} \Bigl({t \over t_0}\Bigr)^{-{1 \over \gamma + 1}} \eqno (53)$$
and writing in (51)
 $$k\ =\   {\gamma + 1 \over \gamma} {c_M \over R_M} \eqno (54) $$
with 
 $$R(t)\ =\   R_M a(t) \eqno (55) $$
where $R_M$ and $c_M$ are constant and $a(t)$ is dimensionless, yields
 $$ds^2\ =\   a^2(t)\ [c^2_M d\vartheta ^2 - R^2_M d \xi ^2] $$
or 
 $$ds^2\ =\   a^2(t)\ [c^2_M d\vartheta ^2 - d \lambda ^2] \eqno (56)$$
with
 $$d \lambda\ =\   R_M d \xi  \eqno (57)$$
to give the spatial distance $d \lambda$ the dimension of a length. Thus $\vartheta$ 
can be identified with the conformal time, which gives the Robertson-Walker
metric an expression of the conformal type, and $c_M$ can be interpreted as the 
constant light velocity in the Minkowski space-time which is measured when using coordinates $\vartheta$ and 
$\lambda$. It may be noted that, up to a scaling factor, a unique definition of the conformal time is
obtained here for the dust universe as well as the for the
radiation one -- which does not happen in the classical theory.\par
 So as to study the formation of the redshift, let us consider, as in [10], 
two events occurring at the same point in space and separated by a time 
interval corresponding to (56):
 $$\Delta s_1\ =\   c_M a(t_1) \Delta \vartheta \quad , \eqno (58)$$
where this time interval is small as compared to time $t_1$.
 If these two events consist in the emission of two light signals which will be
perceived at another point in space, the time interval between the two instants
of reception will be the same in conformal time as in (58), that is, 
$\Delta \vartheta$. This results directly from (56): light propagation is described by
 $$d \lambda\ =\   \pm c_M d \vartheta \eqno (59)$$
so that, for a propagation along the radial coordinate $\chi$,
 $$d \lambda ^2\ =\   R_M^2 d \chi ^2 \quad , \eqno (60)$$
 $$d \vartheta \ =\   \pm {R_M \over c_M} d \chi \quad \eqno (61)$$
hence
 $$\vartheta \ =\   \pm {R_M \over c_M} \chi + {\cal C} \quad , \eqno (62)$$
where $\cal C$ is constant: 
the conformal time needed to go from the emission point to the reception one
depends on the difference of the radial coordinates of these points only and not
on the instant of emission, so both signals will arrive at the (same) reception
point at two instants separated by the same conformal time interval 
$\Delta \vartheta$ as at their
emission. To this $\Delta \vartheta$ corresponds at the reception time $t_2$ an interval
 $$\Delta s_2\ =\   c_M a(t_2) \Delta \vartheta \quad , \eqno(63)$$ 
hence:
 $${\Delta s_2 \over \Delta s_1}\ =\   {a(t_2) \over a(t_1)} \quad , \eqno (64)$$
or, after (55),
 $${\Delta s_2 \over \Delta s_1}\ =\   {R(t_2) \over R(t_1)} \quad . \eqno (65)$$
 This conclusion is valid whatever the type of coordinates used, either $(t, l)$
or $(\vartheta, \lambda)$. \par
 The remainder of the argumentation relies entirely on the definition of proper time 
 $$d \tau\ =\   {1 \over c} ds \eqno (66)$$
in general relativity: \par 
\indent a) with coordinates $(\vartheta, \lambda)$, $c$ is constant and equals $c_M$,
and proper time $d \tau _{\vartheta}$ is defined by:
 $$ds\ =\   c_M d \tau _{\vartheta} \quad , \eqno (67)$$  
so that
 $${\Delta \tau _{\vartheta _2} \over \Delta \tau _{\vartheta _1}}\ =\   {R(t_2) \over R(t_1)} \quad : 
\eqno (68)$$
the redshift is obtained;\par
\indent b) with coordinates $(t, l)$, proper time $d \tau$ is defined by:
 $$ds\ =\   c(t) d \tau \eqno (69) $$
hence
 $${\Delta \tau _2 \over \Delta \tau _1}\ =\   {\Delta s_2 \over \Delta s_1} {c(t_1)
\over c(t_2)}\ =\   {R(t_2) \over R(t_1)} {c(t_1) \over c(t_2)} \eqno (70)$$
and, if $\gamma$ has the same value at $t_1$ and $t_2$,
 $${\Delta \tau _2 \over \Delta \tau _1}\ =\   \Bigl[{R(t_2) \over R(t_1)}\Bigr]^
{\gamma + 1 \over \gamma}\ =\   {t_2 \over t_1} \quad . \eqno (71)$$
 In the last relation, $\Delta \tau_1$ is the period of the radiation emitted at
 $t_1$ and $\Delta \tau_2$ the period of this radiation when received at $t_2$; now, the redshift is defined by the ratio between $\Delta \tau _2$ and the emission 
period $\Delta \tau _{e_2}$ of the same radiation at the reception place at $t_2$ and, as results from (50),
 $${\Delta \tau _{e_2} \over \Delta \tau_1}\ =\   {t_2 \over t_1} \quad , \eqno (72)$$
so that
 $${\Delta \tau _2 \over \Delta \tau _{e_2}}\ =\   1 \quad : \eqno (73)$$
there would be no redshift if cosmic time intervals were accessible to 
measurement. In the present model they are not, and the redshift is obtained
with conformal time intervals (68). This is a logical consequence of the
absence of an observable permanent reference scale for time: $t$ can be regarded as a purely geometrical variable which might be measured from outside the universe, but not
inside it (and, by definition, there is no outside). The same can be said about
the associated space variables $l$, and the actual physical variables are $\vartheta$
and $\lambda$, physical time being the proper time associated with coordinate
time $\vartheta$ through
 $$d \tau _{\vartheta}\ =\   a(t) d\vartheta \quad. \eqno (74)$$ 
\indent Of course, at a given $t$, intervals $\Delta \vartheta$ of conformal time only can be
measured. As concerns cosmic time and space variables, only relations 
established without using non-measurable quantities (such as in (72)) 
are physically meaningful. For example,
horizon calculations can be performed with either type of variables: \par
\indent a) in geometrical coordinates, they involve ratios of quantities taken
at the same time only, 
the event and the particle horizons being defined at time $t_1$ as
 $$\varrho _e\ =\   \int_{t_1}^{\infty}{c(t)dt \over R(t)} \eqno (75)$$  
and
 $$\quad \varrho _p\ =\   \int_0^{t_1}{c(t)dt \over R(t)} \quad ; \eqno (76)$$
now, from (47) and (53),
 $${c(t) \over R(t)} \ \propto \ {1 \over t} \eqno (77)$$
whatever $\gamma$: there is no particle nor event horizon, and no need for inflation. \par
\indent b) the same result is obvious in conformal coordinates $(\vartheta, \lambda)$ (here the lower limit of integration in $\varrho _p$
should be replaced by $- \infty$, since conformal time extends to infinity, as will
be seen hereafter). \par
\vskip10mm
\indent {\bf 4.\ \ Relation between times $t$ and $\vartheta$}\par
\vskip5mm
 Integrating (51) between $t_0$ and $t$, one gets
 $$ \vartheta (t_0, t)\ =\ {1 \over k} \int_{t_0}^t {dt \over t}\ =\ {1 \over k} \ln {t
\over t_0} \eqno (78) $$
for the conformal time interval corresponding to the interval $(t_0, t)$ of
cosmic time. This implies, introducing a third instant $t_1$:
 $$ \vartheta (t_0, t_1)\ =\ {1 \over k} \ln {t_1 \over t_0} \eqno (79) $$
and
 $$ \quad \vartheta (t_1, t)\ =\ {1 \over k} \ln {t \over t_1} \quad , \eqno (80) $$
hence
 $$ \vartheta (t_0, t)\ =\ \vartheta (t_0, t_1) + \vartheta (t_1, t) \quad : \eqno (81) $$
conformal time is additive. \par
 A cosmic time interval can be defined as an ordered pair $(t_1, t_2)$ to be 
associated with the ratio $t_2 / t_1$ (or even $R(t_2) / R(t_1)$) -- as in (71)
 for instance:
 $$ \phi(t_1, t_2)\ =\ {t_2 \over t_1} \quad . \eqno (82) $$
 This defines on cosmic time intervals a multiplicative group law
 $$ \phi(t_0, t_2)\ =\ \phi(t_0, t_1) \phi(t_1, t_2) \eqno (83) $$
isomorphic to the additive one specified by (78) and (81). And indeed, a 
theorem from group theory states that every one - parameter connected 
differentiable group is isomorphic to an additive one and that the additive
parameter is unique [11, 12]: so $\vartheta$ only is additive,
and it can be identified with the linear time $\theta$ introduced by Misner 
[13] as
 $$ \theta \ \sim \ \ln {R(t) \over R(t_0)} \quad . \eqno (84) $$
 Thus, the three possible time notions (cosmic, conformal and 
linear) of L\'evy - Leblond [11] reduce here to two ones only. As already seen,
one of them (conformal time $\vartheta$) can be measured and thus appears to 
coincide with physical time, whereas, by the lack of  
observability of its reference scale, cosmic time $t$ cannot. However (50, 51), $t$ shows 
up as the dilation parameter of intervals of cosmic time, in the same way as
$R(t)$ is the dilation parameter of intervals of length, so that not only
space, but space-time itself, is in expansion. It may be noted that this 
interpretation of $t$ already seems to be possible for Kepler's third law: if
the larger axis of an orbit is multiplied by a factor $R$, the period must be
multiplied by a factor $t$ such that $t^2 / R^3$ is constant, hence the possibility
of interpreting the relation $R \sim t^{2/3}$ as a relation between the scale
parameters for space and time dimensions. Indeed, the third law expresses the 
invariance of Kepler's problem in the inhomogeneous dilation generated by the
infinitesimal operator 
 $$ X\ =\ 3 t {\\ \over \\ t} + 2 x^i {\\ \over \\ x^i} \eqno (85) $$
(N. H. Ibragimov [14]).\par 
One notion of time only being classically used in physics, it seems interesting to 
approximate the relation between $\vartheta$ and $t$ by a linear function, that is to
replace the function $\vartheta (t)$ by its tangent in the neighbourhood of an instant
$t$, thus shifting from $t$ to $\vartheta$ through a change of scale and origin which does
not modify the equations of physics.\par
Substituting $t_1 + \Delta t$ for $t$ in (78), one gets:
 $$ \vartheta \ =\ {1 \over k} \ln {t_1 + \Delta t \over t_0} \quad ,\eqno (86) $$
hence
 $$ \vartheta \ \simeq \ {1 \over k} (\ln {t_1 \over t_0} + {\Delta t \over t_1}) \eqno
 (87) $$
or
 $$ \vartheta \ \simeq \ {1 \over k} (\ln {t_1 \over t_0} + {t \over t_1} -1) \quad . 
\eqno (87') $$
Now, inverting (78),
 $$ t\ =\ t_0 \exp (k \vartheta) \quad , \eqno (88) $$
gives in (47)
 $$ R(t)\ =\ R(t_0) \exp ({\gamma \over \gamma + 1} k \vartheta) \quad ; \eqno (89) $$
this expression  can be approximated using (87) by
 $$ R(t)\ \simeq \ R(t_1) (1 + {\gamma \over \gamma + 1} {\Delta t \over t_1})
 \eqno (90)$$
(which might have been obtained directly from (47)): if $t_1$ is large, (for 
instance corresponding to present time, so that $t_1$ is the "age of the 
universe"),
this approximation is satisfactory even for large values of $\Delta t$ as
soon as they are small relative to $t_1$; if $t_1$ is small, in the 
neighbourhood of 0, the same linear approximation for $\vartheta$, in the form (87'),
yields
 $$ R(t)\ \simeq \ R(t_1) \exp \bigl[ {\gamma \over \gamma + 1} \bigl( {t \over
 t_1} - 1 \bigr) \bigr] \eqno (91) $$ 
or
 $$ R(t)\ \sim \ {\cal C} \exp \bigl( {\gamma \over \gamma + 1} {t \over t_1}
 \bigr) \quad , \eqno (92) $$ 
a result which is valid for $t$ in the vicinity of $t_1$ and which gives, when
extended to larger values of $t$, the same variation of $R$ with cosmic time
as in inflation theory (take for instance $t_1 = 10^{- n}$ s in (92)). \par
 As a last remark, the relation between cosmic and conformal times might help
understand -- if this is confirmed -- why the oldest stars seem to be older 
than the universe itself. It may be noted that, defining $t_M$ by 
 $$ t_M\ =\ {\gamma \over \gamma + 1} {R_M \over c_M} \quad , \eqno (93) $$
(51) can be given the form
 $$ dt\ =\ {t \over t_M} d\vartheta \eqno (94) $$
so that $dt = d\vartheta$ at $t = t_M$: the cosmic and the conformal times are in 
coincidence (have the same scale) for $t = t_M$. Equation (79) for example  
becomes with this notation
 $$ \vartheta (t_0, t_M)\ =\ t_M \ln {t_M \over t_0} \quad . \eqno (95)$$
Now, the age of a star leaving the main sequence of the H-R diagram is 
currently estimated from the time it has spent on this sequence, which is the
longest one in its lifespan. This time is determined by dividing the total
nuclear energy available on the main sequence by the amount of energy used per
time unit, and of course is computed as a conformal time interval. 
The relation between this interval and the cosmic times $t_M$ and $t_0$ which 
respectively represent the age of the universe and the time the star 
entered the main sequence is given by (95) for a star which leaves the main 
sequence presently; it can be rewritten as
 $$ t_0\ =\ t_M \exp (- \vartheta (t_0, t_M) / t_M) \quad . \eqno (96) $$
Applying (96) to the results of Pierce et al [15], which imply $t_M \simeq 7.3
\ 10^9$ years for $\Omega = 1$, gives, with $\vartheta (t_0, t_M) \simeq 16.5\ 10^9$ 
years for the age of the oldest star clusters
 $$ t_0 /t_M \ \simeq \ \exp (-2.26) \ \simeq \ 0.1 \quad , \eqno (97)$$ 
thus reducing to about $7\ 10^8$ years the cosmic time at which these were 
formed.\par
\vskip10mm  
\indent {\bf 5.\ \ Euclidian subspaces in a constant negative curvature manifold}\par
\vskip5mm
 A well-known example of such a situation is that of 3-space in the space-time
of steady-state cosmology [9]. Schr\"odinger [16], who remarked that curvature 
depends on the frame, had already studied such a 
case as the Lema\^\i tre-Robertson frame of the de Sitter universe, represented
as a one-shell hyperboloid ${\cal H}_1$
 $$ x^2 + u^2 + v^2 + y^2 - z^2\ =\ {\cal R}^2 \eqno (98) $$
embedded in a five dimensional space with a (1, 4) Lorentzian metric
 $$ d\sigma ^2\ =\ - dx^2 - du^2 -dv^2 - dy^2 + dz^2 \quad . \eqno (99) $$ 
 What we need here -- so as to have a negative curvature space and not 
space-time -- is a two-shell hyperboloid ${\cal H}_2$
 $$ x^2 + u^2 + v^2 + y^2 - z^2\ =\ - {\cal R}^2 \eqno (100) $$
embedded in the same Lorentz space, with the difference that this imbedding
space is completely fictitious here, so that, in particular, no coordinate
system in it has to be interpreted as including time as one of its 
components.\par
The Lema\^\i tre transformation leading to Euclidean subspaces of ${\cal H}_1$
reads
 $$ \bar x\ =\ { {\cal R} x \over y + z} \quad , \eqno (101)$$
 $$ \bar u\ =\ { {\cal R} u \over y + z} \quad , \eqno (102)$$
 $$ \bar v\ =\ { {\cal R} v \over y + z} \quad , \eqno (103)$$
 $$ \bar \theta \ =\ \ln {y + z \over {\cal R} } \quad . \eqno (104)$$
With these new variables, the $d\sigma ^2$ (99) becomes
 $$ d\sigma ^2\ =\ - \exp(2 \bar \theta) (d \bar x^2 + d \bar u^2 + d \bar v^2) + 
{\cal R}^2 d \bar \theta ^2 \eqno (105) $$
as can be shown using the following relations deduced from (101 -- 104):
 $$ x\ =\ \bar x \exp(\bar \theta)\ , \quad  u\ =\ \bar u \exp(\bar \theta)\ , \quad
 v\ =\ \bar v \exp(\bar \theta) \quad , \eqno (106)$$ 
 $$ y + z\ =\ {\cal R} \exp(\bar \theta) \quad , \eqno (107)$$
and
 $$ y - z\ =\ {\cal R} \exp(- \bar \theta) - {\bar r^2 \over {\cal R}} \exp
(\bar \theta) \eqno (108)$$
where 
 $$ \bar r^2\ =\ \bar x^2 + \bar u^2 + \bar v^2 $$
and $y - z$ has been computed from
 $$ y - z\ =\ {{\cal R}^2 - x^2 - u^2 - v^2 \over y + z} \quad .$$ 
It is clear from (105) that the subspaces $(\bar x, \bar u, \bar v)$ are 
Euclidean.\par
The same Lema\^\i tre transformation leads to Euclidean subspaces of 
${\cal H}_2$ with the only differences that now
 $$ y - z\ =\ {- {\cal R}^2 - x^2 - u^2 - v^2 \over y + z}\ =\ - {\cal R}
\exp(- \bar \theta) - {\bar r^2 \over {\cal R}} \exp(\bar \theta) \eqno (109)$$
and
 $$ d\sigma ^2\ =\ - \exp(2 \bar \theta) (d \bar x^2 + d \bar u^2 + d \bar v^2) - 
{\cal R}^2 d \bar \theta ^2 \quad . \eqno (110)$$
The Euclidean subspaces $(\bar x, \bar u, \bar v)$ of ${\cal H}_1$ 
and ${\cal H}_2$ are the three 
dimensional generalizations of the parabolae traced on their restrictions to
the dimensions $(x, y, z)$ by planes $y + z = {\cal C}$, a constant, and it
may be noted (as concerns ${\cal H}_2$) that when ${\cal R}$ varies with time 
(expansion), $\bar \theta$ remains constant, so that these subspaces are preserved. 
\par
Now, depending on the value of ${\cal C}$, there is an infinity of such 
subspaces and it seems to be logical, so as to keep (as supposed from the 
beginning) a zero four dimensional density, to associate to each subspace with
density $\rho$ another one with negative density $- \rho$ in a twofold 
structure with
 $$ \rho\ >\ 0 \ ,\quad \quad  G\ >\ 0 \quad \quad {\rm on\ the\ 1^{st}\ sheet}$$
and
 $$ \rho\ <\ 0\ ,\quad \quad  G\ <\ 0 \quad \quad {\rm on\ the\ 2^{nd}\ sheet}$$
both of which possibilities being allowed as solutions of (43) (it may be noted
that $\rho G$ also, and not $G$ alone, appears in the Einstein equations).
$\rho$ is an active gravitational mass density, so that two masses should 
attract when in the same sheet and repel when not if one keeps on identifying
passive gravitational mass and inertial mass. Such a negative density sheet has
already been proposed as a candidate for dark matter in first simulations of the large scale structure of the universe [17, 18].\par  
\vskip10mm
\indent {\bf 6.\ \ The case of non-Euclidean subspaces}\par
\vskip5mm
Such subspaces obviously exist (for $k = -1$, take for instance $y^2$ constant
in (100) and for $k = 1$, $z^2$ constant and 
larger than ${\cal R}^2$). However, supposing $k = -1$ in steps (24 -- 30)
implies a zero density (hence the necessity to take $n > 3$),
and $k = 1$ yields $\gamma = 1/2$ and $1$ for the radiation and the matter 
dominated eras respectively, that is no correspondence with previous theories.
Hence the Friedmann solution with $k = 0$ is the only one which can be found 
following the present procedure.\par 
\vskip10mm
\indent {\bf 7.\ \ Gauge relations}\par
\vskip5mm
Since, in geometrical coordinates, $c$ has been assumed to vary with time, this 
should be the same for other physical constants also. This implies gauge
relations which have been proposed in [19, 20] for the case $\gamma = 2$ 
(equation 
(38)). Similar relations can be obtained here, without a special treatment
being required for Schr\"odinger's equation. Assuming every mass varies as $M$
and 
the Bohr radius as $R$, as already supposed, and using (38) yields for 
instance, if $\alpha$ (which is dimensionless) remains constant,
 $$ \hbar \ \approx \ M R^{1 - {1 \over \gamma}} \quad .\eqno (111) $$
In the same way, starting from Newton's formula for the two-body problem 
and still using (38) gives for the Einstein constant
 $$ {G \over c^2}\ \approx {R \over M} \eqno (112)$$
whatever the value of $\gamma$ is. With no precision about the variance of masses, such
relations are purely kinematical -- being derived from the variances of length
and time intervals. One more hypothesis has to be made to go further; for 
example, if $G / c^2$ is to be a constant, this implies 
 $$ M\ \approx \ R \eqno (113)$$
hence
 $$ \hbar \ \approx \ R^{2 - {1 \over \gamma}} \eqno (114) $$
and 
 $$ \quad G \ \approx \ R^{- {2 \over \gamma}} \quad , \eqno (115)$$
and so on, \dots \par
None of these variations can be detected in the laboratory since, as
in the case of length and time intervals, no permanent reference scale can be
used for that. So, in the physical world, constants are but constants, as
light velocity $c_M$ in (56).\par
However, the above assumption about $G / c^2$ is important: it allows the 
divergenceless character of the energy-momentum tensor to be 
kept in Einstein's equations (2'). Moreover, 
substituting (47) into (41) yields for the energy density 
corresponding to the equation of state (39)
 $$\varepsilon \ =\ {3 c^2 \over 8 \pi G} \Bigl({\gamma \over \gamma + 1}\Bigr) 
t_0^{-2} \Bigl({R \over R_0}\Bigr)^{-2 {\gamma + 1 \over \gamma}} \eqno (116)$$
so that a constant value of $G / c^2$ implies
 $$\varepsilon \ \propto \ R^{-2 {\gamma + 1 \over \gamma}} \eqno (117)$$
and for the energy contents of a covolume $V$,
 $$E \ \propto \ R^{1 - {2 \over \gamma}} \eqno (118)$$
so that during the radiation era ($\gamma = 1$),
 $$\varepsilon \ \propto \ R^{-4} \eqno (119)$$ 
and
 $$E \ \propto R^{-1} \eqno (120)$$ 
-- hence, using the Planck formula, the same variation with $R$ of the 
temperature of the radiation gas, as also results from the classical 
treatment -- and during the matter-dominated era ($\gamma = 2$),
 $$\varepsilon \ \propto \ R^{-3} \eqno (121)$$ 
hence a constant energy for the massive particle contents of $V$.\par 
\eject
\indent {\bf 8.\ \ Discussions and conclusion}\par
\vskip5mm
In order to express the hypothesis of non-observability of permanent reference
scales for time and length intervals, we have assumed the scale parameter 
$R(t)$ applies to any scale in the universe -- which implies the possibility of a variation with cosmic time of the constants of physics. Thus a clear distinction is made between the universe and the permanent scales with respect to which its time and space characteristics (its age and its scale parameter) are defined, so that these permanent scales are to be considered as external to the universe.
As opposed to what
can be expected, this does not prevent the redshift from being obtained in the 
resulting theory if one notes that cosmic time (and length) intervals -- for 
which permanent reference scales have to exist as soon as one writes down 
expressions (3 -- 5) or (52) for the metric -- are no longer measurable quantities. The 
physical quantities, measurable from inside the universe, are from this viewpoint the proper time and the proper
distances associated with conformal time $\vartheta$ and comobile coordinates $\lambda$
in equation (56) (as $\vartheta$ only, and its associated proper time, can be
measured, one may wonder why not drop $t$: however, no permanent reference
scale is supposed to exist for $\vartheta$, and if time is to be geometrized,
it seems it should be endowed with this property).
Thus time and space are dealt with in the same way, as contrasts with the
classical theory: as well as space, cosmic 
time expands, and a time arrow arises naturally. The
cosmological and the gravitational redshifts both originate in the variations 
of the $g_{0 0}$ coefficient of the metric, the former with time, the latter
with position in space -- although in both cases, the shift owes its existence
in the phenomenon (the emission of a radiation) being produced and observed as
two distant events in space-time.\par 
In physical (i.e. conformal) time, the origin of the universe
is shifted to $-\infty$, corroborating the ideas developed by L\'evy - Leblond and Misner in [11] and [13], and this is linked with the logarithmic dependence of $\vartheta$ on $t$ as results from (51). This removes the question of knowing what existed before the big bang and might help to solve the problem raised by the age of the oldest star clusters. Other interesting aspects of the model are the absence of horizons and the fact it yields the Euclidean 
character of one of the Friedmann 
models, in full agreement with many observations.\par   
As regards the interpretation, the fact that intervals of $\vartheta$ and $\lambda$ 
only are physically measurable implies the impossibility of directly observing any 
expansion -- hence there is no problem about where it stops; however the metric (56) is not stationary, hence the redshift.
Concerning the matter-dominated era, only experiments can help to decide whether scale
invariance holds: structures such as
atoms or galaxies may well withstand expansion owing to their internal forces; 
as for the radiation era, the absence of horizons is an interesting 
characteristic which may supply possibilities for an alternative to 
the scalar field which is evoked to justify inflation at the microscopic level.
\par  
Interestingly, as results from the 
resolution of the system (29, 30), the density corresponds to the 
curvature of $n$-space; the derivative of $c$
with respect to cosmic time appears in the pressure term only. \par
The possibility of an objection to the model might be found in the 
identification of terms involving 
a variable $c$ with the energy-matter terms of another one (the
Friedmann model) which does not suppose this variation. In fact, it may
be noted that if both sides of (2') are multiplied by $c^2$, so that 
$R_{\mu \nu}$ is replaced everywhere by $c^2 R_{\mu \nu}$, the Ricci tensor 
$c^2 R_{\mu \nu}^{(F;1+3)}$ (equations (24) and (24')) of the zero curvature Friedmann model (and of this 
one only) does not depend on $c$. And the classical equations of this model
write:
 $$2 {R'' \over R} + {R'^2 \over R^2}\ =\ - {8 \pi G \over c^2} p \eqno (122)$$
and
 $$3 {R'^2 \over R^2}\ =\ {8 \pi G \over c^2} \varepsilon \quad , \eqno (123)$$
which can also be derived from (29) and (30) above: they involve $G / c^2$ only
when expressed in terms of the density of energy $\varepsilon$. With 
$G /c^2$ constant as indicated above, there is no 
contradiction in identifying the classical terms with those of the present  
model.\par 
As a last remark, it may be noted that the basic 
hypotheses we have started from involve both general relativity and, through 
the expression of the Bohr radius, a fundamental result of quantum mechanics --
which are both, at an elementary level, connected with an inverse square law. 
\par
\vskip10mm
\indent {\bf Acknowledgements}\par
\vskip5mm
We wish to thank Drs. Guy Oliver and Ivan Brissaud and Profs. Michel Mizony 
and L\'eon Brenig for numerous fruitful discussions.
\vskip10mm
\indent {\bf References}\par
\vskip5mm
\noindent\ [1] L. Nottale: Fractal space-time and microphysics, World 
Scientific, Singapore (1993).\par
\vskip2mm
\noindent\ [2] S. K. Blau: Answer to question \#15 ["What space scales 
participate in cosmic expansion?", F. Munley], Am. J. Phys. {\bf 63}, 779 - 780
(1995).\par
\vskip2mm
\noindent\ [3] C. Callender and R. Weingard: Answer to question \#15 \dots ,
Am. J. Phys. {\bf 63}, 780 (1995).\par
\vskip2mm
\noindent\ [4] P. S. de Laplace: \OE uvres compl\`etes, Gauthier - Villars,
Paris (1878).\par
\vskip2mm
\noindent\ [5] H. Poincar\'e: La valeur de la science, Flammarion, Paris (1973).
\par
\vskip2mm
\noindent\ [6] F. Hoyle: Astronomy and cosmology, Freeman, San Francisco (1975).\par
\vskip2mm
\noindent\ [7] P. Wesson: The properties of matter in Kaluza - Klein cosmology, 
Mod. Phys. Lett. {\bf A 7}, 921 - 926 (1992).\par
\vskip2mm
\noindent\ [8] J. Ponce de Leon and P. S. Wesson: Exact solutions and the 
effective equation of state in Kaluza - Klein theory, J. Math. Phys. {\bf 34},
4080 - 4092 (1993).\par
\vskip2mm
\noindent\ [9] S. Weinberg: Gravitation and cosmology, Wiley, New - York (1972).\par
\vskip2mm
\noindent\ [10] L. D. Landau et E. M. Lifschitz: Th\'eorie du champ, Mir, Moscou
(1962)
\par
\vskip2mm
\noindent\ [11] J. M. L\'evy - Leblond: Did the big - bang begin?, Am. J. Phys.
{\bf 58}, 156 - 159 (1990).\par
\vskip2mm
\noindent\ [12] M. Hamermesh: Group theory, Addison - Wesley, Reading (1964).\par
\vskip2mm
\noindent\ [13] Ch. W. Misner: Absolute zero of time, Phys. Rev. {\bf 186},
1328 - 1333 (1969).\par
\vskip2mm
\noindent\ [14] N. H. Ibragimov: Sophus Lie and harmony in mathematical physics 
on the 150th anniversary of his birth, The Mathematical Intelligencer {\bf 16},
20 - 28 (1994).\par
\vskip2mm
\noindent\ [15] M. J. Pierce, D. L. Welch, R. D. Mc Clure, S. van den Bergh, 
R. Racine and P. B. Stetson: The Hubble constant and Virgo cluster distance
from observation of cepheid variables, Nature {\bf 371}, 385 - 389 (1994).\par
\vskip2mm
\noindent\ [16] E. Schr\"odinger: Expanding universes, Cambridge University Press (1957).\par     
\vskip2mm
\noindent\ [17] J. P. Petit: The missing mass effect, Il Nuovo Cimento 
{\bf B 109}, 697 - 710 (1994).\par
\vskip2mm
\noindent\ [18] J. P. Petit: Twin universe cosmology, Astrophysics and space
science {\bf 226}, 273 - 305 (1995).\par
\vskip2mm
\noindent\ [19] J. P. Petit: An interpretation of cosmological model with
variable light velocity I, Mod. Phys. Lett. {\bf A 3}, 1527 - 1532 (1988).\par
\vskip2mm
\noindent\ [20] J. P. Petit: Cosmological model with variable light velocity: 
the interpretation of red shifts, Mod. Phys. Lett. {\bf A 3}, 1733 - 1744 
(1988).\par   
\end